 \documentclass[final,5p,times,twocolumn,authoryear]{elsarticle}

\usepackage{amssymb}

\newcommand{\apss}{APSS\ }
\newcommand{\aj}{AJ,\ }
\newcommand{\apj}{ApJ,\ }
\newcommand{\apjl}{ApJL,\ }
\newcommand{\apjs}{ApJS,\ }
\newcommand{\araa}{ARAA,\ }
\newcommand{\pasj}{PASJ,\ }
\newcommand{\ssr}{Space Sci. Rev.,\ }
\newcommand{\mnras}{MNRAS,\ }
\newcommand{\aap}{A\&A,\ }
\newcommand{\aaps}{A\&AS,\ }
\newcommand{\sxp}{IGR~J00569$-$7226}
\newcommand{\maxij}{MAXI~J1659$-$152}
\newcommand{\swift}{{\em Swift}}

\journal{Journal of High Energy Astrophysics}

\begin{document}

\begin{frontmatter}

\title{The Galactic Transient Sky with \swift}


\author{Jamie A. Kennea}

\address{The Pennsylvania State University, 525 Davey Lab, University Park, PA 16802, USA}

\begin{abstract}

  The unique capabilities of \swift\ that make it ideal for discovery and
  follow-up of Gamma-Ray bursts also makes it the idea mission for
  discovery and monitoring of X-ray Transients in the Milky Way and the
  Large and Small Magellanic Clouds. The Burst Alert Telescope allows for
  detection of new transient outbursts, the automated follow-up
  capabilities of \swift\ allow for rapid observation and localization of the
  new transient in X-rays and optical/UV bands, and \swift's rapid slewing
  capabilities allows for low-overhead short observations to be obtained,
  opening up the possibility of regular, sensitive, long term monitoring of
  transient outbursts that are not possible with other currently
  operational X-ray missions. In this paper I describe the methods of
  discovery of X-ray transients utilizing \swift's BAT and also
  collaboration with the MAXI telescope. I also detail two examples of
  X-ray transient science enabled by \swift: \swift\ discovery and monitoring
  observations of \maxij, a Black Hole candidate Low Mass X-ray
  Binary in the Galactic Halo, which has the shortest known orbital period
  of any such system; and \swift\ monitoring of \sxp, an edge on
  Be/X-ray binary that displayed a outburst in 2013 and 2014, and which
  monitoring by \swift\ allowed for detection of dips, eclipses and the
  determination of the orbital parameters, utilizing a measurement of
  doppler shifts in the pulsar period.

\end{abstract}


\end{frontmatter}


\section{Introduction}
\label{intro}

The flare-up of an X-ray transient typically signals a rapid increase in
the rate of accretion onto a compact object, a white dwarf (WD), neutron
star (NS) or black hole (BH), and provides an ideal laboratory for studying
astrophysics in a relativistic regime. Even though transients have been
studied for many years, our understanding of the processes behind extreme
accretion events remains relatively poor. The term X-ray transients cover a
wide range of different system phenomenology, but is typically used to mean
Low Mass X-ray Binary (LMXB) and High Mass X-ray Binary (HMXB) systems
containing BH and NS secondaries. However, it may also refer to a wide
range of transient X-ray phenomena including millisecond pulsars
(e.g. \citealt{Campana08}), magnetar outbursts (e.g. \citealt{Kennea13a}),
Stellar Flares (e.g. \citealt{Drake14}) and many others.

The process of accretion that drives most X-ray astrophysical phenomena can
often be dramatic and short lived, with increase in accretion rates causing
X-ray flux rises of up to 6 orders of magnitude, in the case of Supergiant
Fast X-ray Transients (\citealt{Romano14} and references therein), from
quiescent levels. In many cases these events lead to the discovery of
previously unknown systems, or systems that were previously considered
uninteresting. These transient events are rare and often short lived,
making detection and detailed study difficult. Other transients outbursts
may be from sources that were previously known outbursts, but have not been
seen for many years, for example BH transient outbursts are known to be
recurrent, but the time between outbursts has been reported to be as long
as 60 years \citep{Eachus76}. To obtain a good rate of detection of
transient outbursts, X-ray instruments that cover very large areas of the
sky are required.

However, wide field and all-sky instruments, such as {\em Fermi} GBM
\citep{Meegan09}, {\em MAXI} \citep{Matsuoka09}, {\em INTEGRAL} ISGRI
\citep{Lebrun03} and {\em RXTE} ASM \citep{Bradt93}, typically lack the
spatial resolution required to provide accurate localizations necessary for
further optical and IR observations, and typically do not have enough
sensitivity for a detailed analysis of the characteristics of the outburst.

NASA's \swift\ mission \citep{Gehrels04} was designed to localize bright
X-ray transient events, in this case Gamma-Ray Bursts (GRBs). Its three
instruments, the Burst Alert Telescope (BAT; \citealt{Barthelmy05}), the
X-ray Telescope (XRT; \citealt{Burrows05}) and UV/Optical Telescope 
(UVOT; \citealt{Roming05}), provide a unique complement of instruments to
discover and follow-up Gamma-Ray bursts. This combined with a spacecraft
that provides very rapid and accurate slewing to a target, allows for
reporting of Gamma-Ray burst positions within minutes of them being
detected.

The same capabilities that make \swift\ so successful at finding GRBs, are
equally as well tuned for discovery and localization of bright X-ray
Transients. In addition the rapid slewing capabilities of \swift\ allows for
low-overhead short ($\sim1-2$\,ks) observations to be taken, allowing for
long-term sensitive monitoring of outbursts to be performed. No other
operating mission is capable of high cadence monitoring outbursts in this
manner.

In this paper I will explore \swift's ability to detect, localize and
follow-up X-ray transients in the Milky Way and the Large and Small
Magellanic Clouds, as both a stand-alone telescope and in concert with
other X-ray/Gamma-Ray wide field detectors utilizing the \swift\ Target of
Opportunity (ToO) program.




\section{\swift\ discovery and localization of X-ray Transients}

\swift\ performs observations of new X-ray transients utilizing triggers from
both the BAT and from other wide field X-ray and Gamma-ray
observatories. In this section I will describe the various methods of
detection and follow-up that are commonly used to enable X-ray Transient
science with \swift.

\subsection{BAT Triggered X-ray Transients}

The BAT covers approximately 1.4 steradian of the sky at any one time, and
due to \swift's diverse observing strategy, both caused by the large number
of targets observed in a typical day, and by the need observe at least 3
targets per 96 minute orbit in order to avoid looking too close to the
Earth, BAT on average covers 80$-$90\% of the sky daily \citep{Krimm13}.
Such near all-sky coverage means that it is excellent at detecting new
X-ray transients that emit in the 15-150 keV BAT energy range.

When BAT detects a bright unknown transient, it triggers the \swift
``automated target'' (AT) response, which is the same response for GRBs: The
BAT localization of the transient, with an error of typically
$\sim3$\,arc-minutes, is telemetered to the ground through the Tracking and
Data Relay Satellite System (TDRSS), and if possible \swift\ will slew to the
coordinates of the transient, and begin follow-up observations with the
UVOT and XRT instruments. All \swift\ TDRSS telemetered products are
distributed through the Gamma-ray Coordinates Network (GCN;
\citealt{Barthelmy95}), enabling community follow-up of newly discovered
transients.

\begin{figure}
\resizebox{\hsize}{!}{\includegraphics[angle=0]{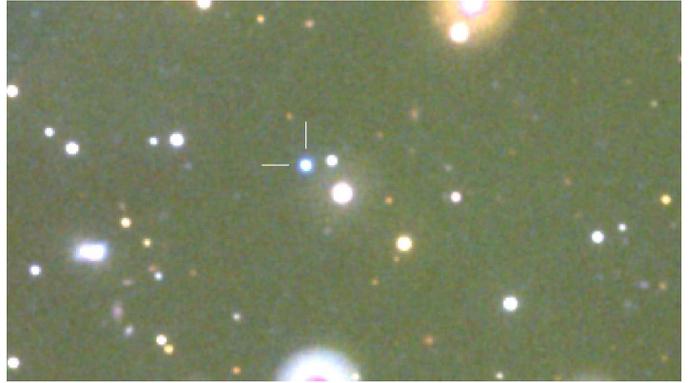}}
\caption{Example of the localization of an X-ray transient with UVOT, in
  this case the transient is the blue star in a cluster of three objects
  near the center of the image.}
\label{uvot}
\end{figure}

When observations begin, the XRT takes a series of short images of the
field ($0.1$\,s and $\sim2.5$\,s long) and attempts to locate the transient
utilizing an onboard centroiding algorithm. If the transient is bright
enough to be detected in this exposure, this location will be telemetered
to the ground through TDRSS within minutes of the initial detection. XRT
will perform observations in Auto state, where the CCD will be read out
in either in Windowed Timing (WT) or Photon Counting (PC) mode based on the
brightness of the new transient (for a description of XRT modes see
\citealt{Hill04}. If PC mode data is taken, event data are telemetered to
the ground through TDRSS for the first orbit, and event reconstruction and
astrometric correction, utilizing UVOT data (e.g. \citealt{Evans09}),
allows for a position to be determined with accuracies up to 1.5
arc-seconds radius, with XRT data alone allowing an accuracy of up to 3.5
arc-seconds radius (all errors quoted at 90\% confidence). UVOT also takes
observations of the field and telemeters these through TDRSS, allowing for
a rapid localization of any optical counterparts of the transient within
minutes of detection (see for example, Figure~\ref{uvot}).

BAT triggered response allows for very rapid reporting of the location of a
new X-ray transient to the community, typically within seconds of detection
through GCN alerts, and approximately 10-25 minutes through GCN
Circulars. In addition for X-ray transients, the \swift\ team will issue a
report on the coordinates, along with a preliminary spectral analysis to
the Astronomers Telegram website (ATel), which is the most common way of
reporting new transients in the X-ray transient community.

\subsection{The \swift/BAT Hard X-ray Transient Monitor}

In addition to transients detected by the BAT triggering algorithm, the The
\swift/BAT Hard X-ray Transient Monitor \citep{Krimm13} is a software based
transient monitor that utilizes BAT data taken during normal
observations. The BAT Transient Monitor has a sensitivity of approximately
5.3 mCrab in a day, allowing for the monitoring of many bright known
sources, as well as the detection of new transients. The primary benefit is
the ability to detect sources down to a much fainter level than needed to
trigger BAT (100$-$200 mCrab, D. Palmer, {\it private communication}),
  meaning slow-rising transients can be detected much earlier than with BAT
  itself.

\begin{figure}
\resizebox{\hsize}{!}{\includegraphics[angle=0]{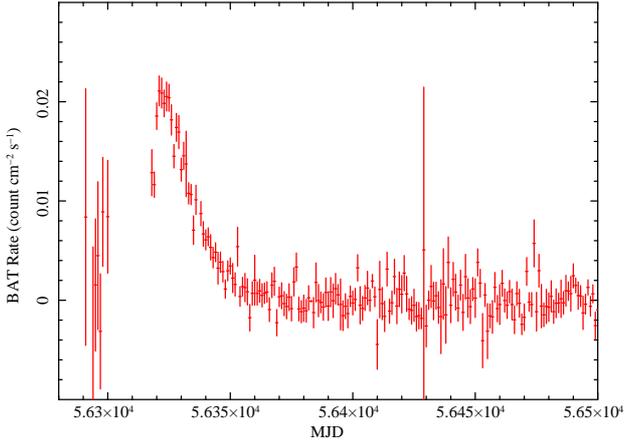} }
\caption{\swift/BAT Hard X-ray Transient Monitor light curve of
 Swift~J1753.7$-$2544 with daily bins. Swift~J1753.7$-$2544 was first detected
 on January 24th, 2013.}
\label{battrans}
\end{figure}

The \swift/BAT Hard X-ray Transient Monitor web site contains light curves
for over 1000 sources, with approximately 250 of these being detected on a
daily basis. Since February 2005 it has discovered $\sim20$ new X-ray
transients. An example of an light-curve of an X-ray transient discovered
by the BAT Transient Monitor, \swift~J1753.7-2544 \citep{Krimm13}, is
shown in Figure~\ref{battrans}.

When a new transient is discovered using this method, observations are
triggered through the \swift\ Target of Opportunity (TOO) program, and
follow-up typically consists of a short PC mode observation in order to
localize the source, follow-up by observations in WT in order to
characterize the spectrum and timing nature of the source. Results of these
observations are reported by the \swift\ team by ATel. Typically these are
sent out within days of the source rising above the detection limit of BAT. 

\subsection{The \swift\ Target of Opportunity Program}

In addition to triggers from BAT, \swift\ performs observations of the new
transients detected by other missions, e.g. {\em INTEGRAL} \citep{Winkler03} and
Fermi/LAT \citep{Atwood09}, through the \swift\ Target of Opportunity
program. The \swift\ TOO program is open to the astronomical community via
the \swift\ Mission Operations Center (MOC) website\footnote{{\tt
    http://www.swift.psu.edu}}, and enjoys a high degree of popularity, for
example, in 2014 951 TOO requests were submitted. 

\swift\ TOOs can be submitted with a variety of priority levels, which
relate to how quickly observations should be performed. Priority 1 (within
4 hours) and Priority 2 (within 24 hours) when submitted will alert the
\swift\ Observatory Duty Scientist (ODS) via mobile phone text message, at
which point they will begin evaluating the feasibility of the requested
observations. Priority 3 (days to a week) and Priority 4 (weeks to a
month), alert the ODS via email.  All TOO requests require approval of the
\swift\ Principle Investigator and/or their deputy before execution.

\swift\ has the capability of performing automated TOO observations through
ground commanding. Given the target’s coordinates, a requested exposure
time and priority level, \swift\ will automatically determine the target
visibility and observe the target for the requested amount of time,
interleaving observations between targets in the current observing plan
based on the prioirty. These TOO commands can be uploaded to the spacecraft
utilizing either TDRSS or ground station commanding. TDRSS commanding
requires manual intervention of the \swift\ Flight Operations Team (FOT), so
is typically only available during working hours, although in some high
priority cases, FOT members will be called into the \swift\ MOC to perform
out-of-hours commanding. 

A customized scheduling system developed by the \swift\ Science Operations
Team (SOT) allows for TOO commanding to be scheduled for a future ground
station pass at any time of the day, allowing for out-of-hours TOO
commanding without the need to call in FOT members. \swift\ typically has ~10
ground station passes per day, meaning that it can be on-target of any TOO
request typically within a few hours at any time of the day.

In addition to single-observation TOOs, \swift\ may observe multiple
locations in order to cover a larger error region than allowable in a
single pointing. There are two ways in which this can be achieved: by
utilizing multiple TOO uploads of differing coordinates in order to 
cover a large error box; and by utilizing the BAT tiling script, which
allows for automated tiling of circular error regions. The former technique
is typically used only for imaging extremely elongated error regions, such
as those from the InterPlanetary Network (IPN; e.g. \citealt{Cline99}).

\begin{figure}
\resizebox{\hsize}{!}{\includegraphics[angle=0]{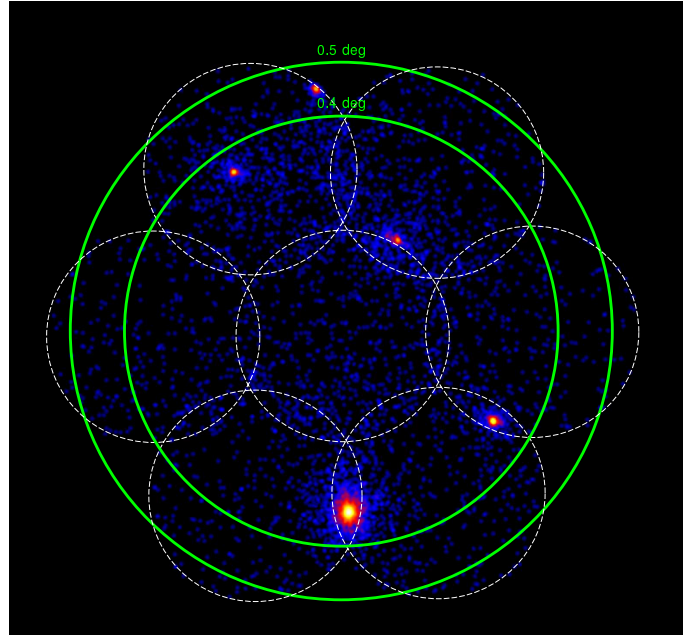}}
\caption{Example of a \swift\ 7-point tiling observation to identify a new
  transient near the Galactic Center. Marked in green are the 0.4 deg error
  circle that gets $\sim100\%$ coverage, and the 0.5 deg error circle that
  gets $\sim95\%$ coverage. Individual XRT pointings are shown as white
  dashed circles. In this field, the brightest X-ray source is A1745-294,
  other sources are all known transients around Sgr A*. }\label{tiling}
\end{figure}

The BAT tiling script allows for tiling of regions utilizing a 4- or
7-point tiling pattern. An example of the 7-point tiling pattern can be
seen in Figure~\ref{tiling}. The 7-point tiling guarantees coverage of an
error circle up to 0.4 degrees radius at $\sim100\%$ coverage, and
$\sim95\%$ coverage of a radius of 0.5 degrees. The benefit of this method
is that as the slews between the individual tiles are very small, they can
be performed quickly and with a low overhead, often allowing for imaging of
the entire error circle in a single \swift\ orbit. Given that Galactic X-ray
Transients are typically bright, observations consisting of seven tiles of
200 seconds exposure each are common. This greatly enhances the ability of
\swift\ to localize transients, even with larger error boxes. As the BAT
tiling script can be executed through the regular TOO commanding, it places
no additional burden on the ODS and SOT, unlike the manual tiling of error
boxes, which would require many more TOO commands to be organized.

\subsection{Monitor of the All-Sky X-ray Image}

Although BAT is very capable at detection of an X-ray Transient outburst,
it’s energy range (15-150 keV) means that is tuned for detection of
transients with relatively hard spectra. This tends to favor BH transients
over NS transients, that often have spectra that break above 10 keV. In
order to increase the number of X-ray Transient detections by \swift, the
\swift\ team partnered with the team of JAXA’s {\em ``Monitor of the All-Sky X-ray
Image''} telescope (MAXI; e.g. \citealt{Matsuoka09}), which is part of the
Japanese Experiment Module on the International Space Station. MAXI is a
scanning telescope consisting of two instruments, the Gas Slit Camera (GSC;
 \citealt{Mihara11})
and Solid-state Slit Camera (SSC; \citealt{Tomida11}). 

Both MAXI instruments are one-dimensional coded mask slit instruments, and
scan majority of the sky every 86 minute ISS orbit. MAXI provides a
powerful tool for the discovery of new X-ray transients, collecting an
X-ray image of the sky in the 0.5-20 keV energy band, with sensitivities as
low as $60$\,mCrab (5 sigma) in a single orbit and $15$\,mCrab in a day. New
transient discoveries are reported by the MAXI Nova Alert system
 \citep{Negoro09} via email communication.

For sources that are relatively stable, MAXI can localize a point source
with an error of approximately 12 arc-minutes radius (90\%
confidence). This matches well the field of view of XRT ($\sim11.8$ arc-minutes
radius), meaning that typically MAXI error circles can be covered with a
single pointing of XRT. Given that MAXI transients are typically $>15$\,mCrab
in brightness, detection by XRT is assured in an a short (500-1000s)
exposure, as XRT’s sensitivity equates to approximately 1 XRT
count/s/mCrab. Fainter and/or temporally variable MAXI transients may have
larger error circles, which require tiling observations to be performed. 

The \swift\ and MAXI teams have set up a collaboration that, through the
\swift\ Guest Investigators Program, follow-up the detection of all new MAXI
detected transients with \swift\ TOO observations in order to confirm and
accurately localize the new transient. These observations are then
typically followed by monitoring observations of this new transient to
track it’s entire outburst.

\section{Selected Transient Results from \swift}

Here I report on the results of \swift\ observations two X-ray
transients. Firstly I cover \maxij, a black hole LMXB in the
Galactic Halo which was discovered by \swift. Secondly I cover 
\swift\ observations of \sxp\  (also known as SXP 5.05), which was
discovered by {\em INTEGRAL}, but which \swift\ observations allowed the discovery
of both eclipses, and measurement of the orbital parameters utilizing
doppler shifts in the pulsar period.

\subsection{\maxij}

\maxij, a BH LMXB system, was first reported after detection by
BAT at 08:05 UT, September 25, 2010. Follow up observations performed by
the XRT and UVOT 31 minutes later localized the transient
 \citep{Mangano10}, although it was initially misidentified as a GRB and
named GRB 100925A. However the GRB identification was disproved when a MAXI
detection $\sim5.5$\,hours before the BAT trigger, confirmed that
\maxij\  was actually a previously unknown Galactic X-ray transient
 \citep{Negoro10}.

UVOT identified a bright ($v=16.7$) candidate inside of the XRT error
circle, this counterpart did not correspond with any known catalog
object. The non-detection of the companion in the USNO-B catalog allowed an
upper limit to be placed on the optical brightness of \maxij\  in
quiescence of $V>21$ \citep{Monet03}, suggesting that the optical
counterpart brightened during outburst by $>4$ magnitudes.

Following the detection by \swift\ and MAXI, a campaign of observations with
a variety of observatories worked to identify the object. IR spectroscopy
confirmed that the optical counterpart showed emission lines consistent
with that of an X-ray Binary \citep{deUP10}. The transient was also
detected in radio \citep{vdH10}, by {\em INTEGRAL} \citep{vovk10}, {\em
 XMM-Newton} \ \citep{Kuulkers10a} and RXTE. RXTE detected a 1.6 Hz type-C
QPO in the power-spectrum, which indicated that \maxij\  was a BHB
 \citep{Kalamkar11}.

A 2.41 hour periodicity, first reported by \cite{Kuulkers10b} from {\em
 XMM-Newton}\ data, and confirmed by \cite{Belloni10} from {\em RXTE} made
\maxij\  the the shortest period BHB yet known. The source of the
periodicity was found to be irregular dips lasting between $5-40$\,min,
with no detecting eclipses \citep{Kuulkers10b}.

As a result of the initial \swift\ trigger, and interest in the source,
\swift\ began a series of daily observations of \maxij, covering the first
27 days of it's outburst, after which the source became too close to the
Sun for \swift\ to observe.

\begin{figure}
\resizebox{\hsize}{!}{\includegraphics[angle=0]{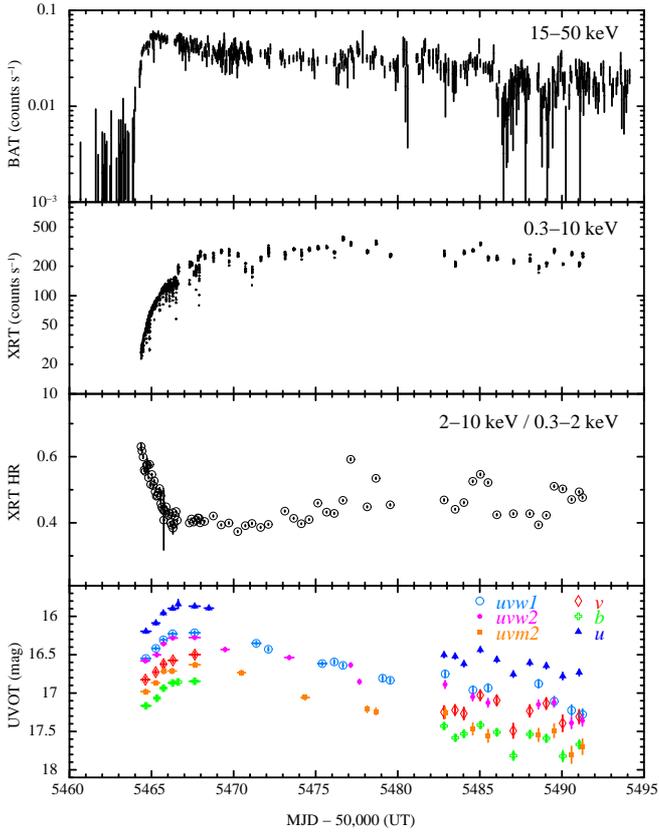} }
\caption{Outburst light-curves of \maxij, from
  \cite{Kennea11}. From top to bottom: $15-50$\,keV BAT Transient Monitor
 light-curve with orbital binning, filtering out bins with less than 500s
 integration times; $0.3-10$\,keV XRT count rate light-curve with 100s
 time bins; Ratio of $2-10$\,keV and $0.3-2$\,keV XRT count rates, binned
 by orbit; UVOT six filter light-curve binned by observing segment. Errors
 are $1\sigma.$}
\label{maxij1659}
\end{figure}

Results of the \swift\ monitoring can be seen in Figure~\ref{maxij1659},
which shows initial outburst light curve of \maxij\  as seen by BAT,
XRT and UVOT. BAT's triggering and \swift's automated slewing capability has
allowed us to see a uniquely detailed and sensitive view of the early part
of the outburst of a BH LMXB Transient. As can be seen in the hardness XRT
ratio plot, there is considerable spectral variability seen in the initial
3-4 days of the outburst. The UVOT counterpart showed variability over all
($v$, $b$, $u$, $uvw1$, $uvw2$, $uvm2$) bands, correlated with the rise
with the X-ray light-curve, peaking with a brightness of $v=16.5$.

\begin{figure}
\resizebox{\hsize}{!}{\includegraphics[angle=0]{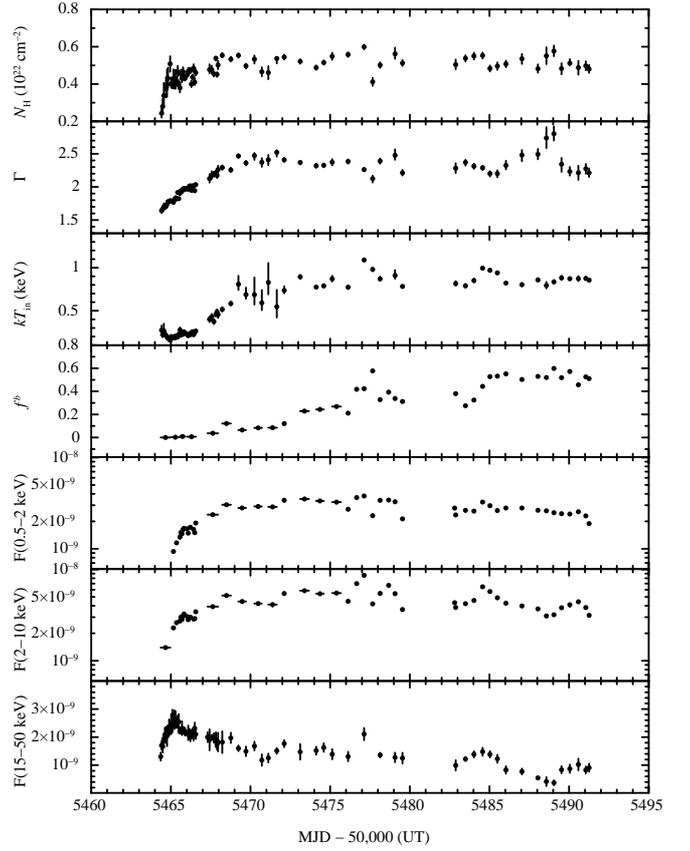} }
\caption{Time resolved spectral fitting of the XRT and BAT data from \maxij, from
  \cite{Kennea11}. Evolution in fitted absorption (top panel), photon index
  (second panel) and inner disk temperature of the XSPEC {\tt diskbb} model
(third panel) are shown. The fourth panel shows the ratio of thermal to
power-law emission, a vital measure of the state of the source, and the
bottom three panels show the flux in 3 energy bands.}
\label{spectralfit}
\end{figure}

By combining XRT and BAT spectra, it was possible to track the spectral
states of \maxij\ (see Figure~\ref{spectralfit}. BHB are known to go through a series of canonical
spectral states (e.g. see \citealt{McandRem06}), which typically can be
defined as the relative combination and strengths of emission from a
thermal disk component, and a comptonized power-law component. In the early
part of the outburst, \maxij\  is dominated by a hard power-law
component ($\Gamma \simeq 1.8$), which quickly steepens
($\Gamma \simeq 2.8$). During the early part of the outburst, the addition
of a significant but small thermal component appears, and grows hotter,
indicating presence of an increasing hot disk component, which inner disk
temperature rises from $\sim.02$ to $\sim0.8$~keV. However, unlike many
other BHBs, \maxij\  never fully transistions into a fully disk
dominated thermal state, remaining instead in an intermediate state, with
the disk never contributing to more than $60\%$ of the total emission in
the 0.5-10 keV XRT energy band.

Given the estimated distance to \maxij\  of $8.6 \pm 3.7$~kpc
 \citep{Kuulkers13}, and the height above the galactic plane
($b^\mathrm{II} = 16.5$), the source lies $2.4\pm1.0$~kpc above the
Galactic plane, suggesting that, similar to other BH binaries such as XTE
J1118+480 \citep{Jonker04}, it is a run-away microquasar that has been
kicked out of the Galactic plane \citep{Yamaoka12}. In fact, \maxij\ appears to belong to a sub-class of short-period galactic BH
LMXBs, along with XTE~J1118$+$480, GRO~J0422$+$32
(e.g. \citealt{Filippenko95}) and Swift~J1753.6$-$0127
(e.g. \citealt{Zurita08}) in the Galactic Halo. 

\swift\ is particuarly suited to study of this subclass of objects, as their
low absorption column allows \swift\ to detect both low absorption X-ray,
important for good quality continuum fitting in XRT's soft X-ray band pass,
and UV/Optical emission, which is typically too extincted to be seen in
Galactic Plane transients.


\subsection{\sxp\  AKA SXP 5.05}

The results discussed in this section have been published by
 \cite{Coe15}. Here I focus on the \swift\ contribution to that paper. 

\sxp\  was first discovered by {\em INTEGRAL} in 2013 in scans of the
Small Magnellanic Cloud performed on October 25-26, 2013 and October 30-31,
2013 \citep{Coe13a}. Follow-up observations were performed by \swift\ on Nov
5th, 2013, observing the field in PC mode for 1-ks. A new X-ray transient
was detected by \swift\ inside the {\em INTEGRAL} reported error circle at the
following coordinates: RA(J2000) = 00h 57m 02.34s, Dec(J2000) = -72d 25m
55.34s with an estimated uncertainty of 2.6 arc-seconds radius (90\%
confidence). This position was consistent with the catalogued massive star
NGC 330-070, a B0.5e type star. This strongly suggested that the \sxp\ was in fact the outburst of a newly discovered Be/X-ray binary
system \citep{Kennea13b}. Archival observations of this star taken in the
second phase of the Optical Gravitational Lensing Experiment (OGLE-II) show
that the system has a likely 17.2 day orbital period \citep{Schmidtke13}

\begin{figure}
\resizebox{\hsize}{!}{\includegraphics[angle=0]{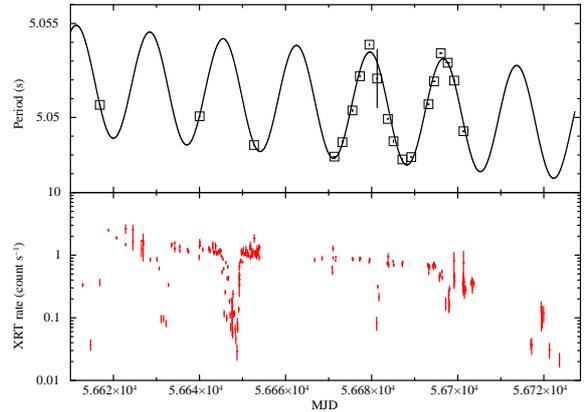} }
\caption{The outburst of \sxp\  as seen by \swift. Top panel: where
possible, the fitted pulsar period utilizing XRT WT data, fitted to these data
are a simple model of a sinusoidal variation, with a linear spin-up. Bottom
panel: The XRT count rate for all observations of the source with XRT,
taken at various cadences.}
\label{sxp505}
\end{figure}

The field containing \sxp\  was observed between 2006 and 2013 for
a total of 25ks. Examining this archival \swift/XRT data showed that the
source was not previously detected and that flux after initial detection by
\swift\ on November 5, 2013, was 3 orders of magnitude brighter than the
upper limit in those observations. 

As a result of the X-ray detection, \swift\ began a series of target of
opportunity observations of \sxp\ . In total \swift\ performed 113
observations of SXP over a period from November 5, 2013 and March 11, 2014,
at which point the transient could no longer be detected by \swift. The
\swift\ light-curve of \sxp\  can be seen in Figure~\ref{sxp505}.

The \swift\ XRT monitoring discovered the presence of broad X-ray dips and
eclipses in the light curve, and combined with {\em INTEGRAL} observations,
confirmed the presence of the 17.2 day orbital period in the X-rays
 \citep{Coe13a}.

\begin{figure}
\resizebox{\hsize}{!}{\includegraphics[angle=0]{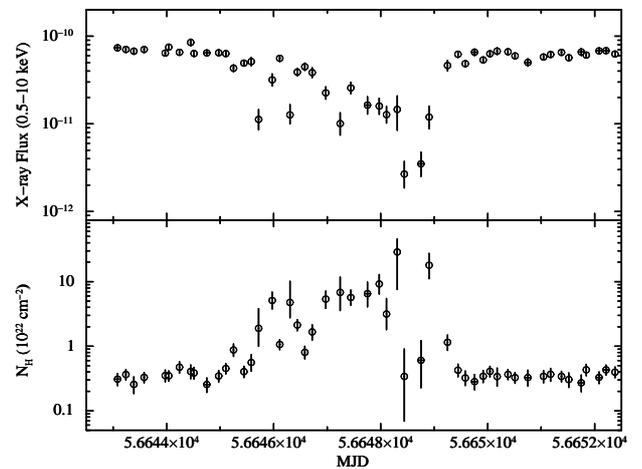} }
\caption{Adapted from \cite{Coe15}, this shows the dip and eclipse present
 in \sxp, and the variation in absorption $N_H$ seen when
 fitting a simple power-law model to the \swift/XRT spectrum. As expected
 with dipping source, the absorption increases with dip depth. During
 eclipse the absorption returns to a value consistent with the out-of-dip
 value, expected if the X-ray seen during eclipse are from the accretion
 disk corona.}
\label{sxp505_eclipse}
\end{figure}

In order to further study the X-ray dips, a series of high cadence
observations were requested in order to cover a predicted dip starting
around December 19th, 2013. The results of these observations can be seen
in Figure~\ref{sxp505_eclipse}, and clearly show not only a broad dipping,
likely caused by attention by the edge of the accretion disk, but also a
short eclipse of the NS by the primary star. Spectral fits of the system
during the dip show the increase of absorption as the dip progresses, with
the $N_H$ value returning to the nominal value, within errors, during the
eclipse, behavior that is consistent with the dip/eclipse interpretation of
this light curve. Given the eclipse \sxp\  must be an near edge on
system, the first edge-on Be/X-ray binary seen in the SMC. The only other
example of an edge on system, LXP 169, is in the LMC \citep{Maggi13}
and does not show eclipses of the NS by the Be/Star. 

In addition to the orbital period, \swift/XRT WT observations of \sxp,
combined with {\em XMM-Newton} observations revealed the presence of an likely
5.05s neutron star spin period. Between January 14, 2014 and March 3rd,
2014 \swift\ peformed a series of 3-5 ks Windowed Timing observations of
\sxp\ approximately every 2 days, after this time period the source had
faded sufficiently that pulsations could no longer be detected in 5ks
observations.

The objective of these observations was to accurately measure the pulsar
period through at least one 17.2 day orbit, in order to measure the doppler
shifts in the pulsar period caused by the orbital motion, with the aim of
finding an orbital solution. Period searching of the WT data was performed
utilizing a $Z^2_2$ test \citep{Buccheri83}, searching with a period resolution
of 1e-5 s, over 10000 steps, producing a period search range of between
5.0-s and 5.1-s. Errors on the measurement of this period were estimated
utilizing the Monte-Carlo method of \citep{Gotthelf99}. Measurements of the
pulsar period in \sxp\  can be seen in the upper panel of
Figure~\ref{sxp505}. Note that these data have been fit with a simple
sinusoidal model combined with a linear spin-up of the pulsar by
accretion. 

As detailed by \cite{Coe15}, these measured periods modeled in detail in
order to find the orbital parameters of the system, and find an orbital
period of $P=17.13 \pm 0.14$ and orbital eccentricity $e = 0.155 \pm
0.018$. This orbital eccentricity is low compared to other Be/X-ray
binaries in the SMC, which typically have $e > 0.3$.

\section{Conclusion}

The \swift\ Mission's unique complement of instruments, that allow it to be a
powerful discovery machine for GRBs, also allow it to be an excellent
observatory for the discovery, localization and follow-up of outbursts from
X-ray Transients. \swift\ enables the rapid localization of transients
discovered by its BAT instrument through automated observation, and the
follow-up of transients discovered by other missions such as {\em MAXI},
{\em INTEGRAL}
and {\em Fermi} through the \swift\ TOO program. It's agile planning and low
overhead observing capabilities allow fast turn around for observations,
allowing the observatory to be on-target within hours of detection. I have
given two examples of transient observations of \swift, which demonstrate
the unique capabilities of the mission: rapid on-target observations, high
cadence monitoring, timing capabilities of XRT, broad band Hard X-ray and
X-ray and UV/Optical follow-up.

\section{Acknowledgments}

This work is supported by NASA grant NAS5-00136. This work made use of data
supplied by the UK Swift Science Data Centre at the University of
Leicester. We acknowledge the use of public data from the \swift\ data
archive. This research has made use of the XRT Data Analysis Software
(XRTDAS) developed under the responsibility of the ASI Science Data Center
(ASDC), Italy.








\end{document}